\newcommand{\Pl}{\partial}
\newcommand{\ts}{\textstyle}

\newcommand{\fpar}[2]{\frac{{\ts \Pl \/ #1}}{{\ts \Pl \/ #2}}}

\newcommand{\bee}{\begin{equation}}
\newcommand{\ene}{\end{equation}}
\newcommand{\beea}{\begin{eqnarray}}
\newcommand{\enea}{\end{eqnarray}}

\newcommand*{\affaddr}[1]{#1} 
\newcommand*{\affmark}[1][*]{\textsuperscript{#1}}
\documentclass[aps,prb,twocolumn,showpacs]{revtex4-1}
\usepackage{graphicx}
\usepackage{dcolumn}
\usepackage{bm}
\begin{document}
\title{Experimental Observation of Electron-Acoustic Wave Propagation in Laboratory Plasma}
\author{Satyajit Chowdhury\affmark[1], Subir Biswas\affmark[2], Nikhil Chakrabarti\affmark[1], Rabindranath Pal\affmark[1].\\
\affaddr{\affmark[1] Saha Institute of Nuclear Physics, 1/AF Bidhannagar, Kolkata 700064, India}\\
\affaddr{\affmark[2] Weizmann Institute of Science, Rehovot 76100, Israel}\\
}
\date{\today}

\begin{abstract}
In the field of fundamental plasma waves, direct observation of electron-acoustic wave (EAW) propagation in laboratory plasmas remains a challenging problem, mainly because of heavy damping. In the MaPLE device, the wave is observed and seen to propagate with phase velocity $\sim1.8$ times the electron thermal velocity. A small amount of cold, drifting electrons, with moderate bulk to cold temperature ratio ($\approx 2 - 3$), is present in the device. It plays a crucial role in reducing the damping. Our calculation reveals that the drift relaxes the stringent condition on the temperature ratio for wave destabilization.  Growth rate becomes positive above a certain drift velocity even if the temperature ratio is moderate.  The observed phase velocity agrees well with the theoretical estimate. Experimental realization of the mode may open up a new avenue in EAW research.
\end{abstract}
\maketitle
\section{Introduction}
In the studies of basic plasma waves and instabilities \cite{krall1973, Bellan2008, pecseli2012}, the linear Vlasov dispersion relation for a uniform, unmagnetized, collisionless plasma produces two well known electrostatic waves which are weakly damped. These are electron plasma wave and ion-acoustic wave (IAW) in high and low frequency regimes, respectively. Both have been established as fundamental plasma modes after extensive studies in laboratory plasmas. Further analysis of the Vlasov equation numerically \cite{Fried1961} revealed another class of solutions in the intermediate frequency range. It is generally known as electron-acoustic wave (EAW) and character-wise  analogous to IAW\cite{monj1}. The experimental observation of EAW along with its propagation characteristics is the subject matter of this article.
The phase velocity of the wave is slightly above the electron thermal velocity. The wave has been studied analytically \cite{Holloway1991, Schamel1986} and numerically \cite{Gary1985} using both kinetic and fluid descriptions of plasmas. In kinetic treatment, it is found to be heavily Landau-damped for a Maxwellian distribution of electron velocity. In the weakly nonlinear limit, trapping of electrons near the phase velocity causes a flattening of the distribution and leads to the mode becoming undamped. In fluid description with single-temperature electron species, the mode does not exist. However, further analysis \cite{Watanabe1977, lakhina2007} showed it to appear if the plasma constitutes of two substantially different electron components, namely bulk hot component (temperature $T_{eh}$) and less dense cold one (temperature $T_{ec}$), along with the neutralizing ion background. In the usual acoustic mode dynamics for this wave, the restoring force of the cold electrons comes from the pressure of the hot component whereas the effective inertia is provided by the cold ones. Numerical analysis \cite{Gary1985} showed EAW to be heavily Landau-damped, unless  $T_{eh}$ $\gg$  $T_{ec}$ - a stringent condition to meet in typical laboratory plasmas.

Importance of EAW lies in the fact that it provides a fast mode of transport for electrostatic disturbances, much faster than the ion-acoustic one. Similar role is played by the Alfven and whistler waves for propagation of magnetic disturbances parallel or near parallel to steady magnetic field, for example in magnetosphere plasma. Furthermore, EAW in nonlinear regime form solitary structures\cite{ikou} and also provides a decay channel for the long wavelength Langmuir waves through nonlinear wave interactions \cite{pecseli1994}. In space plasma, analysis of the broadband electrostatic noise (BEN) in the terrestrial cusp of the magnetosphere and the hiss (high frequency field fluctuations) in the polar cusp region were found to corroborate to the generation of EAW \cite{Thomsen1983, Lin1984, Tokar1984, mace2001}. As mentioned before, in typical laboratory plasmas EAW is highly damped. So its importance has been overlooked for many years and no attempt has been made to study its propagation characteristics. Recently, few laboratory experiments have realized the mode indirectly - for example, spectroscopic phase velocity measurement  from stimulated back-scattered spectra in laser-produced plasma \cite{Montgomery2001}. Flattening of the velocity distribution function by wave trapping is conjectured to be the cause behind its excitation in these experiments. Numerical simulation \cite{Nikolic2002, Ghizzo2006} were done in support of the experiments. Despite being a fundamental plasma mode of significant importance, to our knowledge direct observation of its propagation in laboratory plasma remains an outstanding problem so far. Obstacles for such observation are probably large damping, unsuitable conditions for wave destabilization, very high frequency-very small wavelength for moderate wavenumbers, etc. In this paper we report the direct observation of propagating electron-acoustic waves in very small wavenumber regime after being excited in the plasma of the MaPLE (Magnetized Plasma Linear Experimental) device \cite{Maple2010}. A small amount of drifting cold electrons is present in this plasma. Our analytic treatment reveals these electrons can indeed ease the stringent condition, $T_{eh}$ $\gg$  $T_{ec}$ and destabilizes EAW. The observed dispersion relation matches well with the analytical outcome. Interestingly, in our analysis a critical drift velocity, dependent on density and temperature ratios, is observed above which the mode ceases to exist.

\section{Experimental set up}
\begin{figure}[ht!]
     {\centering
     \resizebox*{8cm}{5cm}
     {\includegraphics[angle=0]{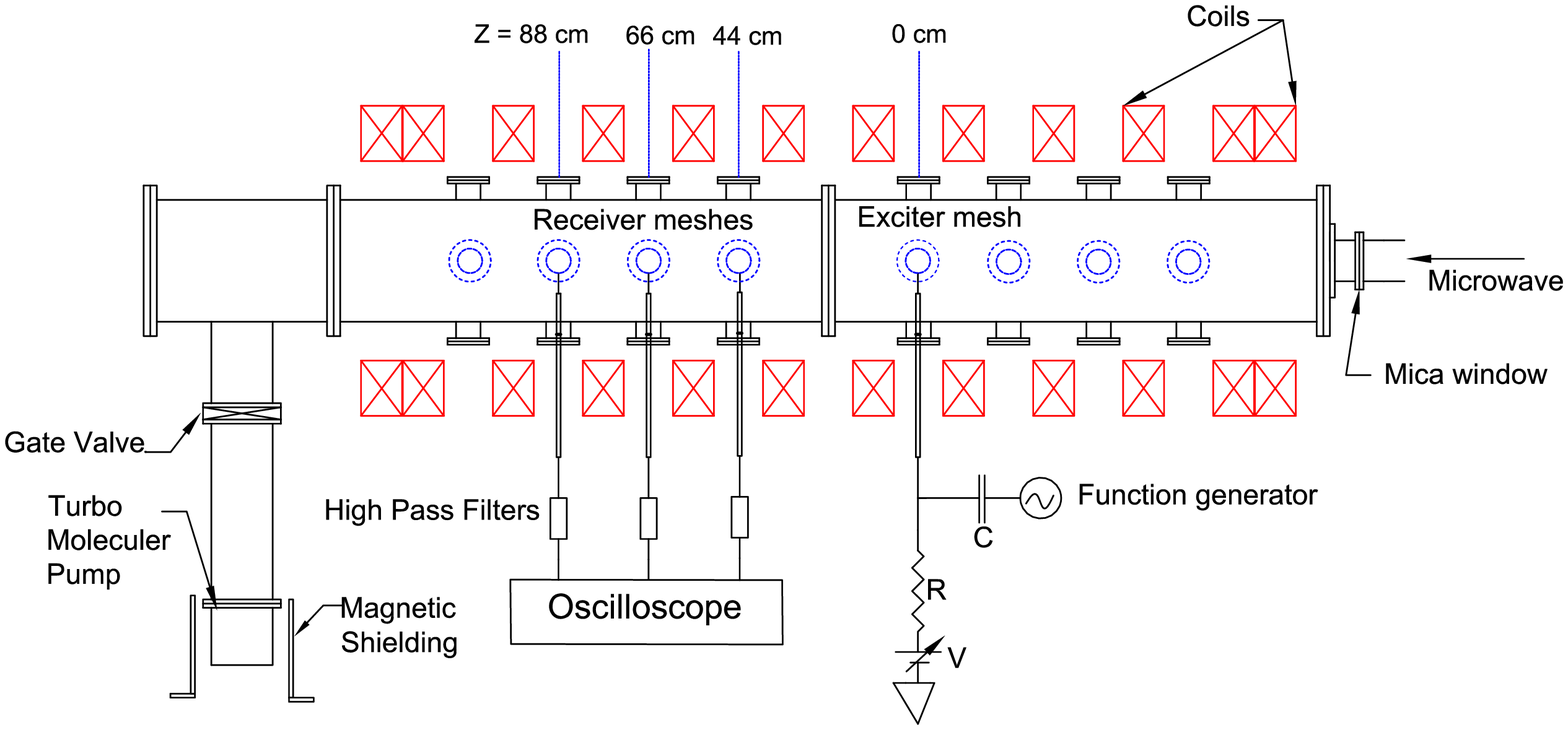}}
     \caption[]{Schematic diagram of the MaPLE device with the experimental setup for wave excitation}.
     \label{fig:maple_schematic}}
\end{figure}

\begin{figure}[ht!]
     {\centering
     \resizebox*{8cm}{5cm}
     {\includegraphics[angle=0]{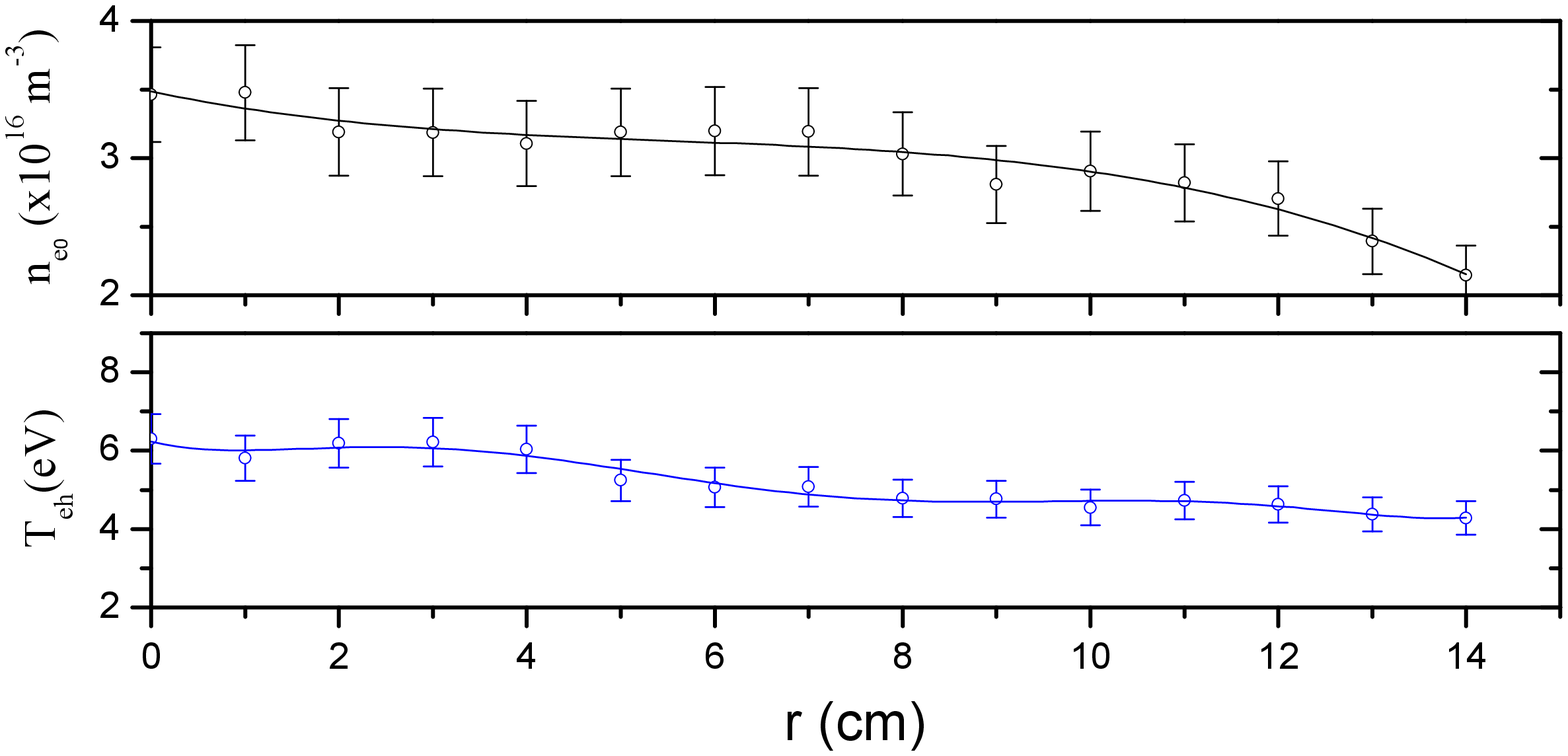}}
     \caption[]{Radial profiles of electron density ($n_{e0}$) and bulk electron temperature ($T_{eh}$). Filling pressure = $5\times10^{-5}$ mbar.}.
     \label{fig:plasma_parameters}}
\end{figure}

The present experiment is carried out in the MaPLE device plasma (diameter 0.30 m and length $\sim 3$ m), produced
by ECR discharge method (Microwave frequency 2.45 GHz, power $\approx 400 W$) in a magnetic field of about 875 gauss. A schematic of the device with the experimental
arrangement is presented in Fig. \ref{fig:maple_schematic}. Detailed description is given
elsewhere \cite{Maple2010}. Nitrogen is the filling gas. Radial profiles of the electron density ($n_{e0}$) and bulk electron temperature ($T_{eh}$) measured with standard Langmuir probe diagnostics at a filling pressure of  $5\times10^{-5}$ mbar are presented in Fig. \ref{fig:plasma_parameters}. The MaPLE device plasma characteristically is known to possess a small amount (a few percent) of colder electron component drifting away from the microwave launching side \cite{Subirtriple2015}. Though the origin of this component is not established yet, but its presence and characteristics are investigated by determining the electron velocity distribution by a retarded field energy analyzer (RFEA) inserted through a radial port. The necessary details of the RFEA can be found in Ref. \cite{Subirtriple2015}. The characteristics are observed to depend on the filling gas pressure. The population density and energy decrease with increasing filling gas pressure and the component disappears at a high filling pressure of ($\sim 5 \times 10^{-4}$ mbar). The component is localized off-axis in the radial region around $r \approx 5$ cm and the drift velocity peaks at $r = 2-3$ cm \cite{Mapleprl}.

The wave launcher is a circular mesh of 4 cm diameter and has about 80 per cent transparency. It is inserted through a radial port located 99 cm from the microwave side end flange. This position is taken as the reference point ($Z=0$) in the axial direction. The vacuum side of the launcher is referred here as `downstream' side where $Z$ is positive, and the microwave side as `upstream' side. The launcher is oriented perpendicular to the magnetic field lines and positioned radially at $r= 5$ cm (Fig.\ref{fig:plasma_parameters}), in the region having drifting cold electron component. Wave is excited by applying a sinusoidal voltage from a function generator to the launcher mesh, which, in addition, is dc-biased. Three identical circular meshes, similar to the launcher, are employed simultaneously to measure floating potentials downstream at axial locations of $Z = 44$, 66, and 88 cm. They too are introduced radially (Fig. \ref{fig:maple_schematic}) and they do not interfere each other as electron Larmor radius and Debye length are very small w.r.t. the mesh hole dimension. Absence of this interference was also confirmed independently by removing the receiver mesh in the middle (placed at Z = 66 cm) and measuring received signal phase difference by the other two (placed at Z = 44 cm and 88 cm). The same time delay was observed in the received signals verifying no interference by the meshes. The amplitude of received signal decreases a little bit in downstream direction. Low frequencies below 1 kHz are first filtered out from the received signals before feeding to a storage oscilloscope (Tektronix DPO 4034). Identical cable lengths are used to eliminate any error in the observed time delays due to cable transmission or filters and have been tested independently by applying test signals.
\section{Experimental results}
\begin{figure}[ht!]
     {\centering
     \resizebox*{8cm}{7cm}
     {\includegraphics[angle=0]{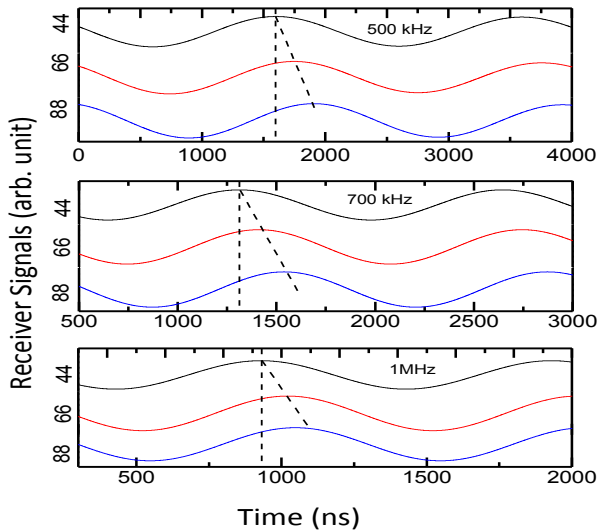}}\par}
     \caption[]{Observation of wave propagation along magnetic field: Received signals at three axial positions Z= 44, 66 and 88 cm for three different exciter frequencies 500 kHz, 700 kHz and 1 MHz. Launcher at $r =$ 5 cm}.
     \label{fig:propagation}
\end{figure}

\begin{figure}[ht!]
     {\centering
     \resizebox*{8cm}{5cm}
     {\includegraphics[angle=0]{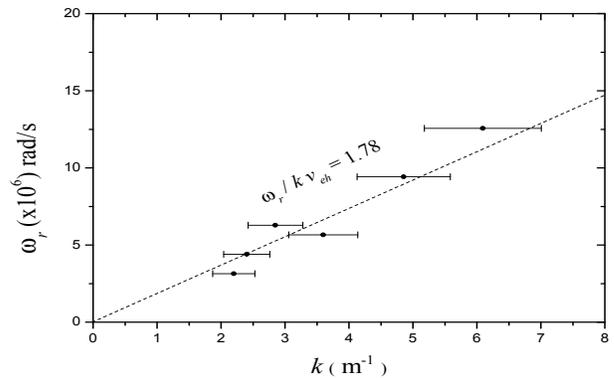}}\par}
     \caption[]{Experimentally obtained dispersion ($\omega$ vs $k$) curve. Dashed line: best fit through the experimental points. ($T_{eh}$= 6 eV)}
     \label{fig:dispersion}
\end{figure}

\begin{figure}[]
      {\centering
  \resizebox*{8cm}{5cm}
     {\includegraphics[angle=0]{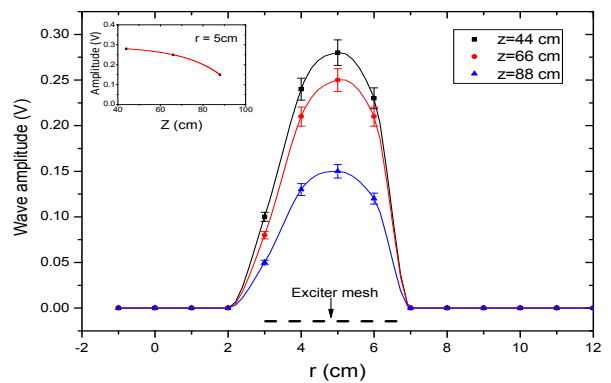}}
     \caption[]{Radial scan of signal amplitudes of the detectors at $Z$ = 44, 66 and 88 cm. Thick line indicates the position of the exciter mesh, $r$ = 5 cm, $Z$ = 0 cm. Inset figure shows the axial variation of signal amplitude at $r$ = 5 cm}
     \label{fig:axial_propagation}}
  \end{figure}

The results of applying a continuous sinusoidal voltage with varying frequencies to the launcher are shown in
Fig. \ref{fig:propagation}, where the
received signals from the three detectors are plotted with time. The launcher is biased negatively with -9 volt and a peak-to-peak driving voltage of 20 volt is applied. It may be mentioned here that the signals are received when the driving voltage goes above a critical value $\sim 10$ volt. The signals are filtered numerically through a narrow band of $\pm 2$ kHz around the driving frequency. The increasing time delays with distance, as seen from Fig. \ref{fig:propagation}, clearly demonstrate the propagation of a wave downstream from the launcher. The phase velocity and wavenumber $k$ can be determined from the time delays and the corresponding receiver distances.
To remove any multiple of $2\pi$ ambiguity, we checked using a sine burst signal on the launcher as a test. We found that phase difference obtained from the burst signal excitation matches exactly with the obtained phase difference in continuous wave excitation. So there is no multiple of $2\pi$ error in the phase difference calculation due to continuous wave excitation. We used the continuous signal as the phase difference is mostly less than $\pi$ to avoid complication in burst signal. Several frequency scans are obtained for verifying the wave propagation. A typical angular frequency $\omega$ vs $k$ data is plotted in Fig. \ref{fig:dispersion}, where $\omega$ is distinctly seen to vary linearly with $k$. The observed phase velocity is $\approx 1.83\times10^6$ m/s. This is much above the
estimated ion-acoustic speed $c_s$ ($=\sqrt{T_{eh}/M}= 6.4\times10^3$ m/s), but slightly above the thermal speed of hot electrons $v_{eh}$ ($=\sqrt{T_{eh}/m}= 1.03\times10^6$ m/s). $M$ and $m$ are ion and electron masses, respectively. Such observations point towards the wave being the electron-acoustic wave that obeys the dispersion relation, $\omega/k$ = (1.78$\pm{0.25}$)$v_{eh}$. The axial propagation of the wave is clearly demonstrated in Fig. \ref{fig:axial_propagation},
where the results of the radial scan of the signal amplitudes observed from all the three receivers ($Z$ = 44, 66 and 88 cm) are presented. The received signal appears only within the projection area of the launcher mesh along the magnetic field line, as we should expect for axial propagation. The wave amplitude is seen to fall slowly as the wave propagates in the axial direction (inset of Fig. \ref{fig:axial_propagation}). This spatial damping may be caused by electron-neutral collision, the rate of which is $\sim 5\times 10^5 s^{-1}$ estimated with the experimental parameters.

\begin{figure}[]
      {\centering
  \resizebox*{8cm}{5cm}
     {\includegraphics[angle=0]{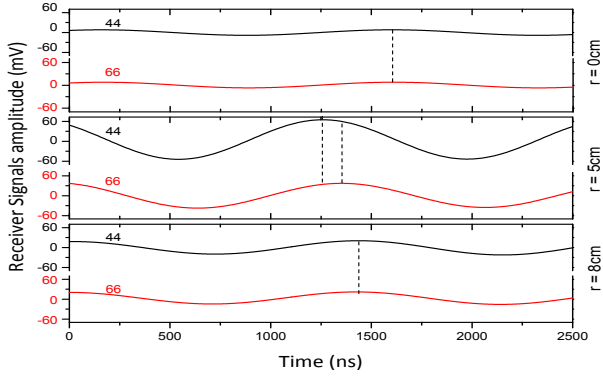}}
     \caption[]{Detected signals at $Z$ = 44 and 66 cm with the launcher placed at three radial locations, $r =$ 3, 5 and 8 cm. Launch frequency 700 kHz.}
     \label{fig:launch_radial}}
  \end{figure}

The exciter mesh is moved radially also to launch the wave at three different radial locations, namely $r =$ 0, 5 and 8 cm. The detected signals at $Z =$ 44 and 66 cm are shown in Fig. \ref{fig:launch_radial}. It is obvious from the figure that the wave is launched at $r=$ 5 cm only; at $r =$ 0 an 8 cm the detected signal levels are very low and also there is no phase shift.

\begin{figure}[ht!]
      {\centering
     \resizebox*{9cm}{7cm}
     {\includegraphics[angle=0]{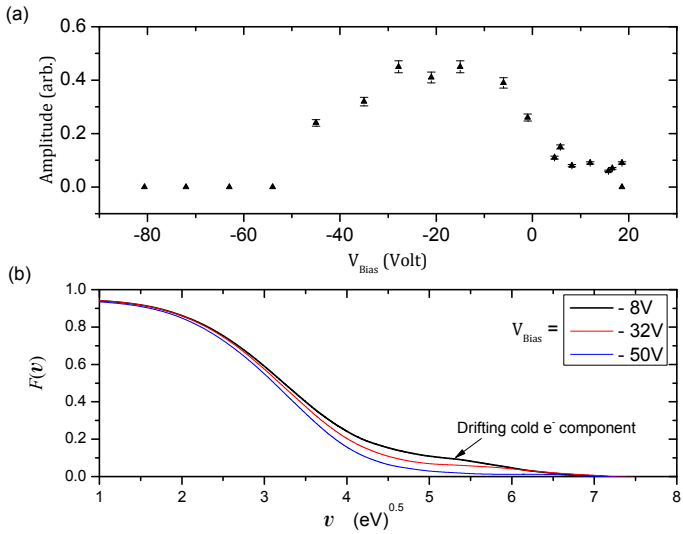}}\par}
     \caption[]{(a) Signal amplitude of the detector at $Z$ = 44 cm as a function of the exciter bias $V_{bias}$, (b) Electron velocity distribution function for different exciter biases. A drifting electron component is seen in the tail. ($Z=44$ cm, $r=5$ cm)}
     \label{fig:signal_with_bias}
  \end{figure}

The bias on the exciter mesh ($V_{Bias}$) is varied and the detected signal at Z= 44 cm receiver is plotted in Fig. \ref{fig:signal_with_bias}a. The wave amplitude is seen to reduce substantially if $V_{Bias}$ is very much negative or positive. It almost disappears for $V_{Bias}$ below -35 volt or above +10 volt. The reasons behind these cut-offs will be discussed later.
\section{Discussions and theoretical analysis}
To investigate the scenario in which the EAW is excited, we first examine the results of the RFEA placed at the position of the first receiver mesh, i.e. at $Z = 44$ cm, $r=5$ cm. The observed electron velocity distribution function with the launcher bias of -9 volt (the same condition as in Figs. \ref{fig:propagation} and \ref{fig:dispersion}) is shown in Fig. \ref{fig:signal_with_bias}b. It has a tail on one side only, as confirmed by orienting the analyzer in opposite direction \cite{Subirtriple2015}. A two-temperature curve with a relative drift seems to fit the distribution nicely. The bulk component represents the hot electrons with $T_{eh}= 6$ eV and the small one the drifting cold electrons with $T_{ec}\approx 3$ eV. The drift energy $V_{c0}$ is 20-25 eV and the density ratio, $n_{eh}/n_{ec}\approx 15$. Fig. \ref{fig:signal_with_bias}b also shows the cold fraction decreases with more negative launcher bias. This is as expected since the sheath thickness around the mesh lines increases with higher value of negative bias. Consequently, the mesh transparency for the drifting cold electrons decreases. Below -40 volt almost no cold electron is present on the downstream side. This explains why no wave is detected in this condition  (Fig. \ref{fig:signal_with_bias}a).

\begin{figure}[ht!]
     {\centering
     \resizebox*{7cm}{4cm}
     {\includegraphics[angle=0]{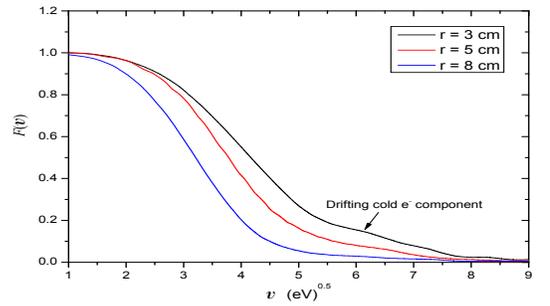}}
     \caption[]{Electron velocity distribution function obtained by RFEA at three radial positions $r =$ 3, 5 and 8 cm. Filling pressure = $5\times10^{-5}$ mbar.}.
     \label{fig:cold_flow_radial}}
\end{figure}

The position of the RFEA is also varied radially, but the analyzer could not be placed at $r=$ 0 cm as high microwave power near the center was disturbing the analyzer. The velocity distribution functions obtained at three radial locations of $r =$ 3, 5 and 8 cm are shown in Fig. \ref{fig:cold_flow_radial}. It is observed that whereas there is a drifting component at 3 and 5 cm, no such component exists at $r =$ 8 cm. Also in a previous experiment \cite{Mapleprl} at low microwave power a mach probe diagnostic showed that there is no drifting component at $r=$ 0 cm. Interestingly, at $r=$ 0 and 8 cm of the launcher position no wave propagation is observed (Fig. \ref{fig:launch_radial}). The temperature of the drifting component though do not vary much, but its relative concentration at 3 cm is about two times higher than that at 5 cm.

The results of Figs. \ref{fig:launch_radial}-\ref{fig:cold_flow_radial} evidently suggest that the cold component has a crucial role in the excitation of EAW. Theoretically, fluid description \cite{Watanabe1977} also
predicts a cold component to be necessary for existence of EAW. However, for low electron temperature ratio such as ours ($T_{eh}/T_{ec} \sim 2$), the numerical analysis \cite{Gary1985} predicts the mode to be strongly damped. No drift was considered in such simulation. A study with drift is therefore necessary to explore whether the damping is reduced so that the electron-acoustic wave can propagate. The problem is basically one-dimensional in nature, so we start with the linear Vlasov dispersion relation for electrostatic modes in one-dimension, given by
\begin{eqnarray}\label{eqn:vlasov}
\varepsilon(\omega,k)
&=&{\left(1-\frac{\omega_{p0}^2}{k^2}P\int_{-\infty}^{\infty}\frac{\fpar{F}{v} dv}{v- \frac{\omega}{k}}\right)}
-i\pi\frac{\omega_{p0}^2}{k^2}\left(\fpar{F}{v}\right)_{{\omega}/{k}}\nonumber\\
&=&0.
\end{eqnarray}
The electron velocity distribution function,
\[F(v)= \frac{1}{\sqrt{2 \pi}\; n_0} \left[\frac{n_{eh}}{ v_{eh}}e^{-v^2/2v_{eh}^2}+\frac{n_{ec}}{ v_{ec}}e^{-(v-v_{c0})^2/2v_{ec}^2}\right]\]
fits our experimental observation. In Eq. (\ref{eqn:vlasov}), $P$ denotes the principal value, $\omega_{p0}(=\sqrt{4\pi n_{0}e^2/m})$ the plasma frequency, $n_{0}(=n_{eh}+n_{ec})$ the total density and $v_{c0} (= \sqrt{2V_{c0}/m})$ the cold electron drift velocity. The ion term is neglected since ions act as an immobile neutralizing background in the time scale of electron response.

With the above distribution function the real part of the dielectric function is approximated as
\begin{eqnarray}\label{eqn:epsilon_real}
\varepsilon_r(\omega,k)=
1-\frac{\omega_{ec}^2}{(\omega-kv_{c0})^2-k^2v_{ec}^2}-\frac{\omega_{eh}^2}{\omega^2-k^2v_{eh}^2}
\end{eqnarray}
Using $\omega=\omega_r+i\gamma$, the equation for the real frequency $\omega_r$ in normalized form comes out to be
\begin{equation}\label{eqn:realfreq}
k^2\lambda_{Dh}^2=\frac{1/\eta}{(x-\beta)^2-1/\alpha}+\frac{1}{x^2-1},
\label{realfreq}
\end{equation}

\begin{figure}[ht!]
      {\centering
     \resizebox*{9cm}{7cm}
     {\includegraphics[angle=0]{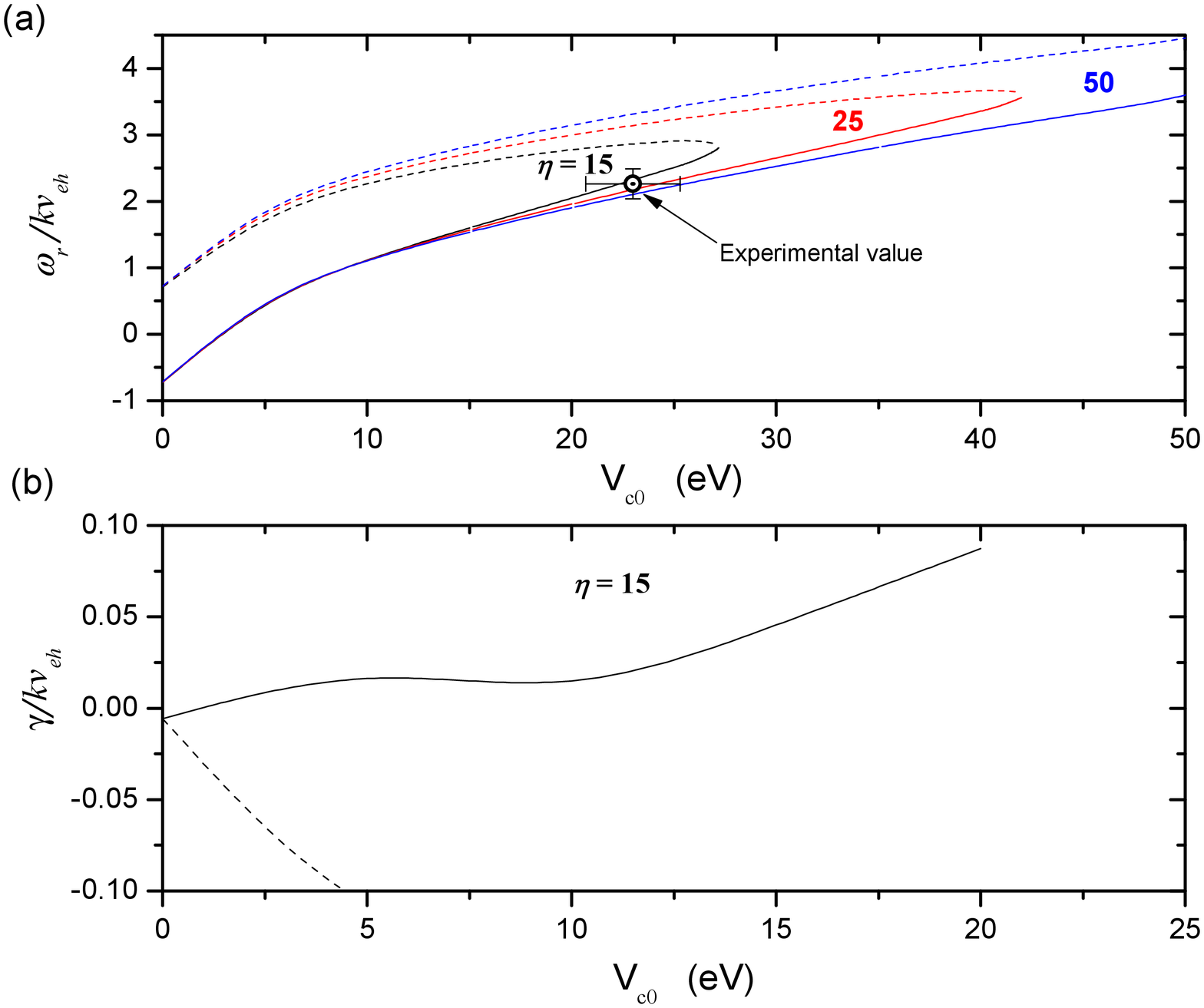}}
     \caption[]{Solution of linear Vlasov dispersion equation: (a) normalized real frequency and (b) growth rate, as function of cold electron drift energy $V_{c0}$ and for three density ratios $\eta$. ($T_{eh}/T_{ec}=2)$. Dashed curves: for forward branch, solid curves: for backward branch. Experimental value of the real frequency is shown with error bars. }
     \label{fig:Vd2expt}}
\end{figure}
where $x=\omega_r/kv_{eh}$, $\alpha= T_{eh}/T_{ec}$, $\eta=n_{eh}/n_{ec}$, $\beta=v_{c0}/v_{eh}$ and the hot electron Debye length $\lambda_{Dh}=\sqrt{T_{eh}/4\pi n_{eh}e^2}$.

Eqn. (\ref{eqn:realfreq}) is solved numerically giving four roots of $\omega_r$. Two roots represent the Langmuir waves, Doppler shifted and commonly known as beam mode. The remaining two belong to the electron-acoustic modes of our interest. In the low wave number regime where $k^2\lambda_{Dh}^2 \ll 1$, the density perturbation of the cold species is balanced by the hot one. It is also the regime of our experimental interest.
To show that we are indeed getting the EAW solution, we take the cold electron equilibrium velocity $v_{c0}=0$. Then, from Eqn. (\ref{eqn:realfreq}) we obtain
                                  \beea
                                    x^2 &=& \frac{\frac{1}{\eta}+\frac{1}{\alpha}}{1+\frac{1}{\eta}} \nonumber \\
                                  \enea
                                               Substituting dimensionless variables
                                      \beea
                                      \omega_r^2 &=& k^2 v_{eh}^2 \left(\frac{\frac{n_{ec}}{n_{eh}}+\frac{T_{ec}}{T_{eh}}}{1+\frac{n_{ec}}{n_{eh}}}\right), \;\;\;\;
                                      \nonumber
                                  \enea
 Furthermore, the simplified EAW dispersion can be obtained as
                                    \bee
                                     \omega_r^2 = k^2 \frac{T_{eh}}{m n_{eh}/n_{ec}}=k^2 c_{se}^2
                                     \ene
 if we assume $T_{ec}/T_{eh} \ll 1$ and $n_{ec}/n_{eh} <1$.
This is the usual non-dispersive electron acoustic wave dispersion relation, well known in laboratory and space plasma community\cite{Gary1985}. Here we must emphasize the physics behind the electron-acoustic wave. It is evident from the dispersion equation that the cold component sustain the wave  and the energy is provided by hot electron temperature. The effective mass of the cold electrons is enhanced by the equilibrium density ratio ($n_{eh}/n_{ec}$). As $n_{eh} > n_{ec}$ in our experiment the mass of the cold electrons is increased and they behave like a heavier species. This mode is basically an analog to ion acoustic wave where cold electrons play the role of ions providing the effective inertia and hot electrons provide the pressure for sustaining the wave \cite{Watanabe1977,monj1}. The solution of the normalized $\omega_r$ from Eqn. (\ref{eqn:realfreq}) is plotted as a function of $V_{c0}$ in Fig. \ref{fig:Vd2expt}a for three values of the density ratio $\eta$. It should be noted that in the frame of the cold species the two modes propagate in opposite directions. We term them as forward (dashed curve) and backward (solid curve) branches. However, in lab frame both of them propagate downstream from the launcher above a critical drift energy. In low wave number regime, the dispersion relation of the modes can be expressed as $\omega_r=akv_{eh}$, where the factor $a$ is dependent on the parameters $\alpha$, $\beta$ and $\eta$.

The growth rate $\gamma$ ($=\frac{\pi\omega_{p0}^2}{k^2}\frac{\partial F}{\partial v}/\frac{\partial \epsilon_r}{\partial \omega}$) is plotted in Fig. \ref{fig:Vd2expt}b using the experimental parameters. With no drift velocity both the EAW-branches are damped ($\gamma < 0$) as expected, but above a critical velocity $\gamma$ becomes positive for the backward branch even if the temperature ratio is low.  Monotonically decreasing distribution, in general, implies no unstable modes. But, the backward branch ($\omega/k < v_{c0}$) belongs to negative energy waves, which occur when the equilibrium has a flow velocity and a mode exits to reduce the average kinetic energy below the equilibrium value. The mode taps the free energy of the flow for its growth. This condition makes $\partial \varepsilon_r/\partial \omega$ negative and $\gamma$ positive. The Landau damping here operates in reverse. It can be visualized by moving to the cold frame where the hot distribution appears as an offset Gaussian. If the offset is large ($v_{c0} > \omega/k$), then $\partial F/\partial v > 0$, giving more fast than slow particles at the wave-phase velocity. This is the mode that can be excited and what we observe in our MaPLE device. The forward branch ($\omega/k > v_{c0}$), on the other hand, remains heavily damped and hence hard to excite. The drift seems to relax the stringent condition on the required temperature ratio for excitation of EAW. A similar kind of relaxation of relatively stringent condition on electron drift velocity by parallel velocity shear of ion flow has been reported in IAW excitation \cite{ganguly,agrimson2001}. In Fig. \ref{fig:Vd2expt}a the average value of $\omega_r/kv_{eh}$ obtained in our experiment from several frequency scans is superimposed. The error bars shown arise mainly from uncertainties in determining $T_{eh}$ and $V_{c0}$. The agreement with the numerical result is reasonably good.

From Fig. \ref{fig:Vd2expt}a it is seen that above a critical drift energy of the cold electrons no solution for pure EAW exists.  This happens because the density perturbations of the two species do not balance each other.With cold fractions going progressively smaller, the critical energy goes higher and higher until the situation arises when $k^2\lambda_{Dh}^2$ term can not be neglected and the mode looses the pure acoustic ($\omega$ proportional to $k$) nature. Above this value arises the beam-like modes in plasma frequency regime - much higher than those observed in this experiment. From the figure we note that this energy is $\sim30$ eV for the present experimental condition ($\eta\approx 15$), which explains why we observe an upper cut off of the launcher bias voltage $V_{Bias}$ (Fig. \ref{fig:signal_with_bias}a) for observing EAW. When the bias goes to high positive value, the drifting cold electrons from the microwave side may be getting accelerated to higher energy and going above the critical value. In our experiment the tail distribution is flat to begin with and EAW travels in one direction. Collisions, being relatively small and neglected in our calculations, can lower the growth rate. This could be the reason behind the observed high driving voltage.
\section{summary}
In summary, we have observed the propagation of electron-acoustic wave directly in the very low wavenumber regime in a laboratory device having two electron species. For moderate temperature ratio of the species the Landau damping should dominate. However, our theoretical analysis shows if the cold component drifts above a critical velocity, the wave can be destabilized. The drift velocity  has an upper limit above which the wave does not exist. The observed experimental results including the phase velocity agree very well with the analysis.
We believe that these experimental findings together with theoretical model could be extended further including more physical facts to provide a new territory for EAW  research as one of the most fundamental plasma mode.
\section{acknowledgement}
The authors acknowledge technical helps from Dipankar Das, Subhasis Basu and Monobir Chattopadhyay. The financial support by Department of Science and Technology of Govt. of India (Project no. SB/S2/HEP-005/2014) is also gratefully acknowledged.
%
\end{document}